\documentclass[a4paper,prl,twocolumn,preprintnumbers,floatfix,superscriptaddress]{revtex4-2}
\usepackage[english]{babel}
\PassOptionsToPackage{utf8}{inputenc}
\usepackage{inputenc}
\PassOptionsToPackage{T1}{fontenc}
\usepackage[super]{nth}
\usepackage{fontenc}
\usepackage{graphicx}
\usepackage{color}
\usepackage[separate-uncertainty=false]{siunitx}
\usepackage{amsmath}
\usepackage{mathtools}
\usepackage{braket}
\usepackage{hyperref}
\usepackage{import}
\usepackage{multirow}
\usepackage{booktabs} 

\makeatletter
\@addtoreset{equation}{supequation}
\makeatother

\makeatletter
\@addtoreset{figure}{supfigure}
\makeatother
\usepackage{xspace}
\newcommand{\fm}[1]{\ifmmode#1\else$#1$\fi}

\newcommand{\ca}{\fm{{}^{40}\mathrm{Ca}^+}\xspace}

\newcommand{\srneutral}{\fm{{}^{87}\mathrm{Sr}}\xspace}

\newcommand{\groundstate}{\fm{{}^2\mathrm{S}_{1/2}}\xspace}
\newcommand{\excitedstate}{\fm{{}^2\mathrm{D}_{5/2}}\xspace}
\newcommand{\clocktrans}{\fm{{}^2\mathrm{S}_{1/2} \leftrightarrow {}^2\mathrm{D}_{5/2}}\xspace}

\newcommand{\BSB}[1]{\fm{\mathrm{BSB_{#1}}}\xspace}
\newcommand{\RSB}[1]{\fm{\mathrm{RSB_{#1}}}\xspace}

\newcommand{\PhiBellsym}{\fm{\ket{\Phi_{\textrm{Bell}}}}\xspace}
\newcommand{\PsiBellasym}{\fm{\ket{\Psi_{\textrm{Bell}}}}\xspace}
\newcommand{\PhiBellsymbra}{\fm{\bra{\Phi_{\textrm{Bell}}}}\xspace}

\newcommand{\ThetaEvolution}{\fm{\Theta}\xspace}
\newcommand{\Thetaro}{\fm{\Theta_\mathrm{ro}}\xspace}
\newcommand{\Thetarel}{\fm{\Theta_\mathrm{rel}}\xspace}
\newcommand{\psicc}{\fm{\psi_{\text{cc}}}\xspace}
\newcommand{\Tint}{\fm{T_{int}}\xspace}
\newcommand{\GammaBell}{\fm{\Gamma_\textrm{Bell}}\xspace}

\begin{document}
	\title{Entanglement-enhanced optical ion clock}
	
	\author{Kai Dietze}
	\affiliation{Physikalisch-Technische Bundesanstalt, Bundesallee 100, 38116 Braunschweig, Germany}
	\affiliation{Institut für Quantenoptik, Leibniz Universität Hannover, Welfengarten 1, 30167 Hannover, Germany}
	\author{Lennart Pelzer}
	\affiliation{Physikalisch-Technische Bundesanstalt, Bundesallee 100, 38116 Braunschweig, Germany}
	\author{Ludwig Krinner}
	\affiliation{Physikalisch-Technische Bundesanstalt, Bundesallee 100, 38116 Braunschweig, Germany}
	\affiliation{Institut für Quantenoptik, Leibniz Universität Hannover, Welfengarten 1, 30167 Hannover, Germany}
	\author{Fabian Dawel}
	\affiliation{Physikalisch-Technische Bundesanstalt, Bundesallee 100, 38116 Braunschweig, Germany}
	\affiliation{Institut für Quantenoptik, Leibniz Universität Hannover, Welfengarten 1, 30167 Hannover, Germany}
	\author{Johannes Kramer}
	\affiliation{Physikalisch-Technische Bundesanstalt, Bundesallee 100, 38116 Braunschweig, Germany}
	\affiliation{Institut für Quantenoptik, Leibniz Universität Hannover, Welfengarten 1, 30167 Hannover, Germany}
	\author{Nicolas C. H. Spethmann}
	\affiliation{Physikalisch-Technische Bundesanstalt, Bundesallee 100, 38116 Braunschweig, Germany}
	\author{Timm Kielinski}
	\affiliation{Institut für Theoretische Physik, Leibniz Universität Hannover, Appelstraße 2, 30167 Hannover}	
	\author{Klemens Hammerer}
	\affiliation{Institut für Theoretische Physik, Leibniz Universität Hannover, Appelstraße 2, 30167 Hannover}
	
	\author{Kilian Stahl}
	\affiliation{Physikalisch-Technische Bundesanstalt, Bundesallee 100, 38116 Braunschweig, Germany}
	
	\author{Joshua Klose}
	\affiliation{Physikalisch-Technische Bundesanstalt, Bundesallee 100, 38116 Braunschweig, Germany}
	
	\author{Sören Dörscher}
	\affiliation{Physikalisch-Technische Bundesanstalt, Bundesallee 100, 38116 Braunschweig, Germany}
	
	\author{Christian Lisdat}
	\affiliation{Physikalisch-Technische Bundesanstalt, Bundesallee 100, 38116 Braunschweig, Germany}
	
	\author{Erik Benkler}
	\affiliation{Physikalisch-Technische Bundesanstalt, Bundesallee 100, 38116 Braunschweig, Germany}
	
	\author{Piet O. Schmidt}
	\affiliation{Physikalisch-Technische Bundesanstalt, Bundesallee 100, 38116 Braunschweig, Germany}
	\affiliation{Institut für Quantenoptik, Leibniz Universität Hannover, Welfengarten 1, 30167 Hannover, Germany}
	
	\date{\today}
	
	\begin{abstract}
		Entangled states hold the promise of improving the precision and accuracy of quantum sensors. We experimentally demonstrate that spectroscopy of an optical clock transition using entangled states can outperform its classical counterpart. Two \ca ions are entangled in a quantum state with vanishing first-order magnetic field sensitivity, extending the coherence time of the atoms and enabling near lifetime-limited probe times of up to \SI{550}{\milli\second}. In our protocol, entangled ions reach the same instability as uncorrelated ions, but at half the probe time, enabling faster cycle times of the clock. We run two entangled \ca ions as an optical clock and compare its frequency instability with a \srneutral lattice clock. The instability of the entangled ion clock is below a clock operated with classically correlated states for all probe times. We observe instabilities below the theoretically expected quantum projection noise limit of two uncorrelated ions for interrogation times below \SI{100}{\milli\second}. The lowest fractional frequency instability of $7 \times 10^{-16} /\sqrt{\tau/\SI{1}{\second}}$ is reached for \SI{250}{\milli\second} probe time, limited by residual phase noise of the probe laser. This represents the lowest instability reported to date for a \ca ion clock.
	\end{abstract}
	\maketitle
	
	\textit{Introduction.\textemdash}
	Since the seminal proposal to use quantum-correlated squeezed states to overcome the shot noise limit in optical interferometry for gravitational wave detectors \cite{caves_quantum-mechanical_1981}, many more quantum sensing applications have been proposed to benefit from using entangled states \cite{pezze_quantum_2018, giovannetti_quantum-enhanced_2004, ye_essay_2024}. Among them are optical clocks, which are the most accurate measurement devices available today, reaching systematic uncertainties at the $10^{-18}$ level \cite{marshall_high-stability_2025, huntemann_2016, Lu_clock_comparison_zhang_barrett_2023, aeppli_clock_2024, zeng_toward_2023, beloy_frequency_2021, mcgrew_atomic_2018, hausser_115in-172yb_2025, takamoto_test_2020, dorscher_optical_2021}. Metrological gain has recently been demonstrated using entangled neutral atoms in optical lattices and optical tweezers \cite{yu_spin-squeezing-enhanced_2024, robinson_direct_2024, cao_multi-qubit_2024, yang_clock_2025}. Ion clocks are typically limited by quantum projection noise (QPN) \cite{itano_quantum_1993}, but increasing the number of ions poses a significant technological challenge \cite{akerman_2025, tan_2019, pelzer_2024_cdd, schulte_multi_ion_2016, arnold_prospects_2015, herschbach_linear_2012, hausser_115in-172yb_2025}. Improving the signal-to-noise ratio using entangled states is therefore particularly attractive for ion clocks \cite{wineland_squeezed_1994, schulte_prospects_2020, bollinger_optimal_1996}. Towards this goal, two-ion Rabi and 14-ion Ramsey spectroscopy on an optical transition using entangled ions \cite{shaniv_toward_2018, monz_14-qubit_2011} have been demonstrated, optimal quantum states for optical clock operation have been implemented on a programmable quantum computer \cite{marciniak_optimal_2022}, and a distributed two-ion entangled state has been employed for optical clock operation \cite{nichol_elementary_2022}. However, a quantum gain in optical clock comparisons exceeding the performance of a competitive version of the clock operating with uncorrelated atoms has not yet been demonstrated. Besides suppressing projection noise, correlated many-particle states can be used to tailor additional desirable features of atomic clocks. In particular, carefully designed entanglement can realize a decoherence-free subspace (DFS) and mitigate systematic or coherence-limiting effects \cite{roos_designer_2006, nichol_elementary_2022}.
	
	In this letter, we report the implementation of an entanglement-enhanced optical \ca{} ion clock with reduced projection noise, employing a generalized Ramsey protocol \cite{optimal_frequency_measurements_bollinger} with a two-ion Bell state encoded in a DFS immune to magnetic field fluctuations. We demonstrate lifetime-limited coherence time in the DFS and clock instability beyond the best demonstrated \ca clock \cite{hao_stability_2024} by comparing to a Sr optical lattice clock \cite{Doerscher_2021}. 
	
\begin{figure}[t!]
		\centering
		\includegraphics[]{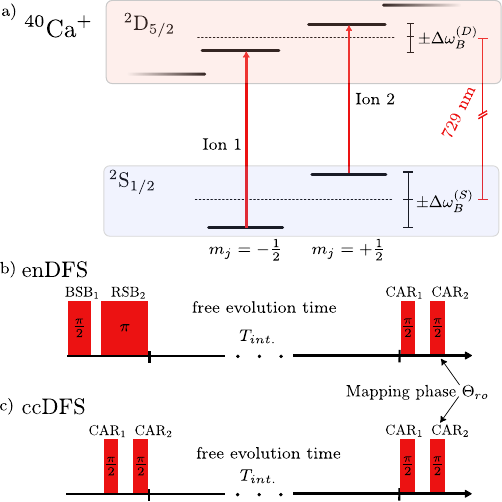}
		\caption{a) Partial level scheme of the \clocktrans transition in \ca. The interferometer is initialized on two different $\Delta m = 0$ transitions.
			Due to different g-factors in the \groundstate and \excitedstate state, the Zeeman-shifts $\Delta \omega_{B}^{(S)}$ and $\Delta \omega_{B}^{(D)}$ result in distinct resonance frequencies for all transitions.
			b) Implementation of the entangled Ramsey-Interferometer (enDFS). Sideband pulses are employed to prepare the state $\PhiBellsym$ (see main text for details).
			c) Implementation of the classically correlated Ramsey-Interferometer (ccDFS). The ions are independently prepared by two $\frac{\pi}{2}$-carrier-pulses. The readout sequence is identical to (b).
		}
		\label{fig:measurement_scheme}
	\end{figure}

	\textit{Entanglement-enhanced protocol.\textemdash}
	We implement a Cirac-Zoller-type entangling gate \cite{cirac_quantum_1995, roos_bell_2004} using resonant interactions between two different pairs of Zeeman states $\ket{S_\pm}=\ket{S, m_j = {\pm 1/2}} \leftrightarrow \ket{D_\pm}=\ket{D, m_j = {\pm 1/2}}$, where $m_J$ is the magnetic quantum number, on the \ca \clocktrans clock transition at \SI{729}{\nano\meter}. A partial level scheme is shown in Fig.~ \ref{fig:measurement_scheme}a). 
	For the entangled DFS (enDFS) scheme, we start from the state $\ket{S_-}\ket{S_+}=\ket{S_-S_+}$ to prepare the Bell state 
	\begin{equation}\label{eq:bell_state}
		\PhiBellsym = \frac{1}{\sqrt{2}} \left( \ket{S_-S_+} + e^{-i\ThetaEvolution}\ket{D_-D_+} \right).
	\end{equation}
	
	The phase evolution $\ThetaEvolution(t)=(\omega_-+\omega_+)t$ of the two-ion wave function during time $t$ is given by the energy difference between the two parts of the wavefunction, where $\omega_\pm$ denote the respective transition frequencies between $\ket{S_\pm}\leftrightarrow \ket{D_\pm}$. Since each part of the wavefunction consists of states that experience an equal, but opposite linear Zeeman shift, the overall state \PhiBellsym is free of this shift \cite{roos_designer_2006}.
	
	During a Ramsey spectroscopy sequence with interrogation time \Tint, and in the limit where the duration of the excitation pulses can be neglected, the relative phase evolution $\Thetarel(\Tint)$ of the spectroscopy laser with frequency $\omega_L(t)$ and the state \PhiBellsym is given by
	\[
	\Thetarel(\Tint) = \int_0^{\Tint} 2\omega_L(t) - \left( \omega_- + \omega_+ \right) \,d t \,.
	\] 
	
	In the readout stage this phase is mapped onto multi-ion states of different parity using resonant $\pi/2$-pulses on both atoms \cite{chwalla_precision_2007}.
	A parity measurement $\hat{P} = \hat{\sigma}_z^{(1)} \hat{\sigma}_z^{(2)}$ of the resulting quantum state has an expectation value $\braket{\hat{P}} = \cos \left( \Thetarel(\Tint) + \Thetaro \right)$, where $\hat{\sigma}_z^{(i)}$ is the Paul spin operator $z$ for ion $i$. The phase $\Thetaro$ contains all relevant laser phases and can be scanned to obtain the contrast of the signal (see Supplemental Material \cite{supplement}), or obtain a feedback signal for laser stabilization.
	
	A direct comparison between the performance of the Bell state and a non-entangled two-ion state for an optical clock is complicated by the fact that the simplest product state $(\ket{S_+}+\ket{D_+})^{\otimes 2}$ is highly magnetic field sensitive and therefore does not allow for the same long Ramsey times as the Bell state. However, starting from the state $\ket{S_-S_+}$, one can also obtain a classically correlated state in the DFS (ccDFS) using a simplified version of the entangled protocol \cite{chwalla_precision_2007}.
	
	Two independent Ramsey $\pi/2$ pulses on the two ions result in the state
	\begin{align}\label{eq:corr_state}
		\ket{\psicc} ={}& \frac{1}{2} \left(\ket{S_-} + e^{-i\omega_- t}\ket{D_-} \right)\otimes\left(\ket{S_+} + e^{-i\omega_+ t}\ket{D_+} \right) \notag \\
		={}& \frac{1}{\sqrt{2}} \left( \PhiBellsym + \PsiBellasym \right), \text{with} \notag \\
		\ket{\Psi_{\textrm{Bell}}} ={}& \frac{1}{2} \left(\ket{S_-D_+} + e^{-i(\omega_--\omega_+)t}\ket{S_+D_-} \notag \right).
	\end{align}
	The state is split into two parts with equal probability. The first part consists of the desired Bell state, \PhiBellsym, in the DFS. The second part, \PsiBellasym, evolves at the difference frequency between the two parts of its wavefunction, which is dominated by their magnetic field splitting. Consequently, this component dephases for Ramsey times exceeding the coherence time of magnetic-field-sensitive states, resulting in a mixed state described by the density operator \(\rho = \frac{1}{2} \PhiBellsym \PhiBellsymbra + \frac{1}{4} \ket{S_-D_+}\bra{S_-D_+} + \frac{1}{4} \ket{D_-S_+}\bra{D_-S_+}.\) While this state exhibits a component of the Bell state $\PhiBellsym$ the density operator overall represents a separable, that is classically correlated, state only. A parity measurement using this density operator will therefore result in the same phase evolution as using the entangled state \PhiBellsym, but with half the measurement amplitude \cite{chwalla_precision_2007, nichol_elementary_2022}.
	
\begin{figure}[t!]
		\centering
		\includegraphics[]{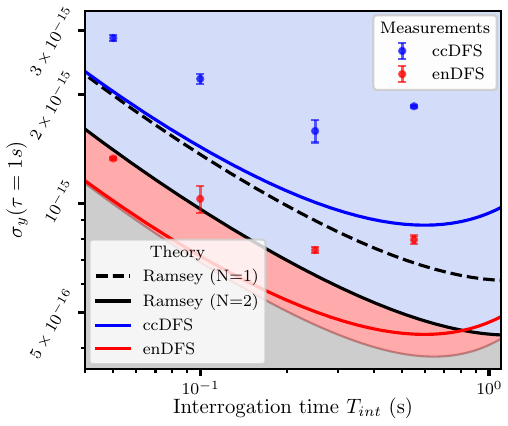}
		\caption{Overview of measured stabilities and theoretical quantum projection noise limits as a function of interrogation time for the three interrogation schemes discussed (enDFS, ccDFS, and standard Ramsey). Theoretical values assume  ideal signal contrast and perfect laser stability, but include a \SI{12.5}{\milli\second} dead time and a reference stability of \SI{1.1e-16}{}$\cdot(\tau/\SI{1}{\second})^{-1/2}$, based on the performance of the $^{87}\mathrm{Sr}$ lattice clock. The coloured regions classify different theoretical stability regimes in the presence of spontaneous decay. Classical interrogation protocols are shown in blue, entanglement-assisted protocols with spontaneous emission error mitigation in red \cite{kielinski_ghz_decay_2024}.}
		\label{fig:stability_figures}
	\end{figure}

	Spontaneous decay of the excited clock state will decrease the measurement contrast for interrogation times \Tint approaching the lifetime of the excited state $t_\mathrm{sp} = \frac{1}{\Gamma}$, where $\Gamma$ is the decay rate from the excited state. 
	Incorporating the spontaneous decay for standard Ramsey spectroscopy with $N$ uncorrelated ions on a transition with frequency \(\omega_0\)  yields a fractional frequency instability after an integration time $\tau$ of 
	\[
	\sigma_y(\tau) = \frac{e^{\Gamma \Tint}}{\omega_0 \sqrt{N}} \frac{1}{\sqrt{\Tint \tau}} \,,
	\]
	which is the QPN limit for uncorrelated atoms.
	The lowest instability is achieved for the optimal Ramsey time $\Tint = 1/\Gamma$ \cite{Peik_2006, Leroux_2017}. Two-particle Bell states exhibit a reduced lifetime $1/\GammaBell=1/(N\Gamma)$, which accelerates the decay of measurement contrast. Consequently, the optimal interrogation time is given by $T_{int} = 1/\GammaBell$, resulting in the same instability as for $N$ uncorrelated particles, but at $N$-times shorter optimal Ramsey times \cite{huelga_improvement_1997} (see also Fig.~\ref{fig:stability_figures}). Well below the lifetime limit, an entangled two-ion clock can reach a gain of $\sqrt{2}$ in stability compared to uncorrelated ions for fixed interrogation times. 
	
	\textit{Experimental implementation.\textemdash}
	A detailed description of the experimental apparatus is given in \cite{hannig_towards_2019} and the Supplementary Material \cite{supplement}. In brief, we trap two \ca ions in a linear Paul-Trap with single-ion trap frequencies $(\omega_{z}, \omega_{x}, \omega_{y})=2\pi\times(0.8, 1.8, 1.9)$\,MHz. A stabilized magnetic field of \SI{0.356}{\milli\tesla} aligned with the axial trap axis lifts the degeneracy of the S- and D-state Zeeman sublevels, spectrally separating the two used transitions by \SI{4}{\mega\hertz}. Residual magnetic field fluctuations broaden the linewidth of these transitions to approximately \SI{100}{\hertz}.
	
	Before the interferometry sequence, the ions are cooled via Doppler- and successive electro-magnetic transparency (EIT) cooling \cite{roos_experimental_2000, scharnhorst_experimental_2018}. The axial out-of phase mode at a frequency of \SI{1.4}{\mega\hertz} is used for the entangling operation and therefore further cooled by resolved sideband cooling close to the ground state of motion. The initial state $\ket{\psi_i} = \ket{S_-S_+}$ is prepared starting with both ions optically pumped into the $\ket{S_-S_-}$ state and transferring one of the ions to the $\ket{S_+}$ state via the $\ket{D_+}$ state using single-ion addressing and a composite pulse sequence (see Supplemental Material \cite{supplement}).

	Cooling, state preparation, detection and clock feedback contribute to a total dead time of \SI{12.5}{\milli\second} per experimental clock cycle. We use red (RSB) and  blue (BSB) sideband pulses which, respectively, add and remove one quantum of motion while driving the ion from S to D during the entangling operations. The Bell state $\PhiBellsym$ for the enDFS scheme is prepared by a $\pi/2$ $\BSB{1}$ pulse on ion 1,  followed by a $\pi$ $\RSB{2}$ pulse on ion 2 (see. Fig.~\ref{fig:measurement_scheme}b)) \cite{roos_bell_2004}. The ions are selectively addressed using a global laser beam, exploiting their frequency separation. The acquired phase $\Thetarel(\Tint)$ is mapped by $\frac{\pi}{2}$-carrier pulses on both ions (CAR$_{1,2}$) onto different parity states. The clock laser is derived from an extended cavity diode laser (ECDL) locked to a high-finesse cavity; its filtered transmission seeds a second ECDL, whose amplified output is used in the experiment \cite{akerman_universal_2015, Nazarova_2008}. The laser frequency is further stabilized via a transfer oscillator lock \cite{scharnhorst_high-bandwidth_2015} to an ultra-stable reference laser using an optical frequency comb. This reference laser \cite{matei_si_cavity_2017} has a coherence time longer than the excited state lifetime of the atoms of $t_\mathrm{sp}\approx \SI{1.17}{\second}$.
	
\begin{figure}[t!]
		\includegraphics[]{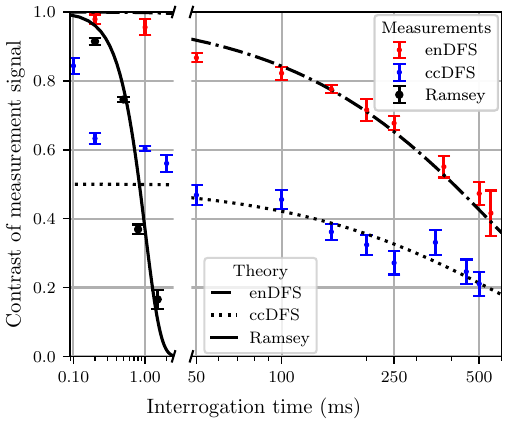}
		\caption{Coherence time for different probe schemes. Contrast for the two-ion entangled (enDFS, red) and correlated scheme (ccDFS, blue) as well as a Ramsey measurement with a single ion (black). The contrast was inferred from fringe-measurements of parity (DFS schemes) and excitation (Ramsey) signal for different interrogation times and fitting a sine function with varying amplitude. The theoretical lifetime-limited contrast is shown for the enDFS (dash-dotted) and ccDFS (dotted) measurement schemes. For short interrogation times, the contrast of the single ion drops in good agreement with a gaussian noise model with \SI{1}{\milli\second} time constant. In contrast, interrogation within the DFS ensures close to lifetime-limited coherence.
		}
		\label{fig:coherence_times}
	\end{figure}

	The coherence time is typically limited by the first-order Zeeman shift, which is $\approx$ \SI{2}{\mega\hertz} at a sensitivity of \SI{5.6}{\mega\hertz\per\milli\tesla} for the two transitions used within this experiment. We determine the coherence time for the enDFS and ccDFS schemes from the contrast of parity oscillations at several interrogation times by scanning the mapping phase. Similarly, for the single-ion Ramsey scheme, we extract the coherence time from the oscillations in the excitation signal. Figure~\ref{fig:coherence_times} shows that the contrast for a single ion is well described by a gaussian noise process $\exp{(-t^2/t_c^2)}$ with a coherence time $t_c\sim\SI{1}{\milli\second}$. In contrast, the effective transition frequency of the DFS state has a strongly suppressed first order Zeeman shift extending the coherence time towards the lifetime limit with a contrast given by the envelope $\exp{(-t\GammaBell)}$.
	A measurement contrast of 0.98/0.46 for the enDFS/ccDFS schemes could be achieved, limited by high frequency laser noise, magnetic field fluctuations and cross-talk between the excitation pulses. The measurements show that the DFS extends the possible interrogation times by more than two orders of magnitude.
	
\begin{figure}[t!]
		\includegraphics[]{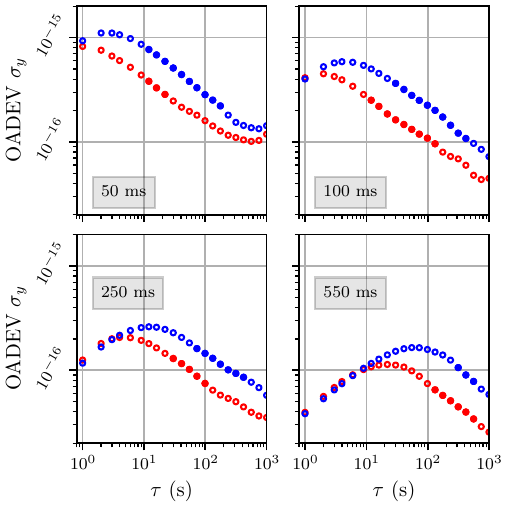}
		\caption{
			Fractional stability of the frequency ratio of the \ca/\srneutral frequency ratio. Data for the enDFS (red) and ccDFS (blue) measurement schemes are shown for four different interrogation times. The fractional stability is given as the overlapping Allan deviation of the data traces (see text for details). Error bars are smaller than the symbol size, filled markers indicate the data range taken for the stability fit.}
		\label{fig:adev_overview}
	\end{figure}

	\textit{Clock stability.\textemdash}
	We operate an optical frequency reference using the enDFS and ccDFS protocols and compare it to a \srneutral lattice clock \cite{Doerscher_2021}, which is well suited for a high stability frequency comparison. All fibers involved in the clock comparison are length stabilized \cite{ye_delivery_2003}. The clock laser is stabilized to the parity signal using a two-point sampling method analogous to \cite{Peik_2006}. During the frequency comparison the contrast is on average 0.92 (0.43) for the enDFS (ccDFS) scheme due to drifts of the motional and carrier frequencies as well as the diminishing fidelity of the single-ion addressing. The contrast directly effects the achievable frequency stability. 
	Fig.~\ref{fig:adev_overview} shows the stability of the frequency ratios for interrogation times of \SIlist{0.05; 0.1; 0.25; 0.55}{\second}, determined using the overlapping Allan deviation (OADEV) \cite{benkler_relation_2015} over averaging intervals ranging from \SI{1}{\second} to \SI{1000}{\second}.
	
	An OADEV as low as \SI{2.5e-17} for \SI{550}{\milli\second} probe and \SI{1000}{\second} averaging time has been achieved for the enDFS scheme. As theoretically predicted, the enDFS scheme shows significantly shorter servo time constants \cite{Leroux_2017, huelga_improvement_1997} and an improvement in stability compared to the ccDFS scheme for all probe times. 
	The instabilities are constrained for long averaging times by a flicker floor that depends on the interrogation time.
	The flicker floor level is compatible with a fluctuating differential phase evolution between the two ions when opening and closing the interferometer: consecutive excitation of the two ions results in short (\SI{20}{\micro\second}) periods of additional phase evolution in magnetic-field-sensitive states. A small and varying detuning ($\approx$\SI{100}{\hertz}) from magnetic field drifts and power line noise accounts for the measured frequency deviations. Especially for the enDFS scheme additional frequency instability over long timescales may arise from ac-Stark shifts of the global laser. These effects can be suppressed using advanced probe schemes \cite{haffner_precision_2003, yudin_hyper-ramsey_2010, huntemann_generalized_2012, sanner_autobalanced_2018, zanon-willette_composite_2018} or mitigated by simultaneous excitation of both ions \cite{Bruzewicz2019}.
	
	Instabilities of the frequency comparison $\sigma_{y,1\mathrm{s}}$ were obtained by a least-squares fit $ \sigma_y(\tau) = \sigma_{y, 1s} \cdot \frac{1}{\sqrt{\tau}}$ to the white frequency noise-dominated regime of the OADEV \cite{freq_stab_ana_riley_2008} (see Supplemental Material for details \cite{supplement}).
	
	We summarize the measured instabilities and compare them with theoretical predictions in Fig.~\ref{fig:stability_figures}. The theory curves have been calculated including the change in duty cycle for varying probe times, while additional noise sources, such as the Dick effect, laser phase noise, and reduced signal-to-noise ratio due to imperfect state preparation and readout, are neglected. This results in an offset of the experimental data from the theoretical predictions.
	As shown in Fig.~\ref{fig:stability_figures}, the theoretical instability of the enDFS scheme is always smaller compared to the ccDFS and uncorrelated two-ion scheme (assuming a perfectly stable magnetic field) for probe times shorter than the optimum probe time. The lowest theoretical instability is identical to a classical two-ion Ramsey scheme, but reached at half the optimal probe time \cite{huelga_improvement_1997}. The ccDFS scheme has an identical scaling to the enDFS scheme, but is shifted to twice the instability due to the smaller signal contrast. The single-ion curve is larger by a factor of $\sqrt{2}$ compared to the two-ion Ramsey scheme. The experimentally observed increase in instability for an interrogation time of \SI{550}{\milli\second} is compatible with uncontrolled optical path length fluctuations of the clock laser beam through the optical setup. 
	
	Experimentally, the enDFS scheme shows an instability below the QPN for two uncorrelated ions at an interrogation time of \SI{50}{\milli\second}. The lowest instability of \SI{7.2e-16} at \SI{1}{\second} is achieved for a probe time of \SI{250}{\milli\second} and is to our knowledge the most stable optical frequency demonstrated with \ca ions \cite{Shi_2024_sr_ca, Hao_2024_ca_clock, Matsubara_2012_ca_sr} and only 45\,\% above the theoretical limit.
	
	\textit{Conclusion.\textemdash}
	We have demonstrated that the magnetic field insensitivity of entangled GHZ in a DFS enables near lifetime-limited probe times and therefore record-low instability. For clocks limited by spontaneous emission from the excited state, a slight variation of the protocol demonstrated here turns out to provide the optimal improvement beyond the standard quantum limit approaching fundamental limits also for larger ensembles up to several tens of atoms (see red-shaded area below the enDFS-limit in Fig.~\ref{fig:stability_figures}) ~\cite{kielinski_ghz_decay_2024}.
	While in the demonstrated schemes the ultimate instability is the same for GHZ states and uncorrelated atoms, the optimum probe time for GHZ states scales with $1/N$. This supports high-bandwidth feedback schemes, which ease the requirements for long-term clock laser stability and systematic shifts such as \nth{2} order Doppler shifts from back ground heating of the ions in the trap \cite{rosenband_frequency_2008, huntemann_single-ion_2016}. Furthermore, the higher locking bandwidth enables searches for new ultralight scalar bosons, such as dark matter candidates, with larger masses corresponding to larger Compton frequencies \cite{arvanitaki_searching_2015, filzinger_improved_2023, beloy_frequency_2021}.
	
	The presented system can be turned into an optical clock by evaluating all systematic shifts. Scaling to more than two ions is complicated by an inhomogeneous electric quadrupole shift of the excited clock state across the ion crystal \cite{itano_external-field_2000, arnold_prospects_2015}. These shifts can be mitigated by employing dynamical decoupling schemes \cite{martinez-lahuerta_quadrupole_2024, pelzer_2024_cdd, akerman_operating_2025, kaewuam_hyperfine_2020, aharon_robust_2019, shaniv_quadrupole_2019, lange_coherent_2020} or more complex multi-ion states in a DFS, enabling multi-ion entangled clocks that perform better than an optimized classical clock of the same species.
	
	\section{Acknowledgments}
	We thank Daniele Nicolodi, Thomas Legero and Uwe Sterr for providing the ultra-stable Si cavity as a laser reference.
	The project was supported by the Physikalisch-Technische Bundesanstalt, the Max-Planck Society, the Max-Planck–Riken–PTB–Center for Time, Constants and Fundamental Symmetries, the State of Lower Saxony, Hannover, Germany through Niedersächsisches Vorab, and the Deutsche Forschungsgemeinschaft (DFG, German Research Foundation), Project-ID 274200144, SFB 1227 DQ-\textit{mat} (projects A06, A09, B02 and B03), and under Germany’s Excellence Strategy – EXC-2123 QuantumFrontiers – 390837967. The project 23FUN03 HIOC has received funding from the European Partnership on Metrology, co-financed from the European Union's Horizon Europe Research and Innovation Programme and by the Participating States. This project has received funding from the European Research Council (ERC) under the European Union’s Horizon 2020 research and innovation program (grant agreement No 101019987).
	\bibliographystyle{apsrev4-2}
	\bibliography{citations_clean}
	
	\section{appendix}
	
	\subsection{Experimental implementation of the measurements}
\begin{figure*}
		\centering
		\includegraphics[width=0.8\linewidth]{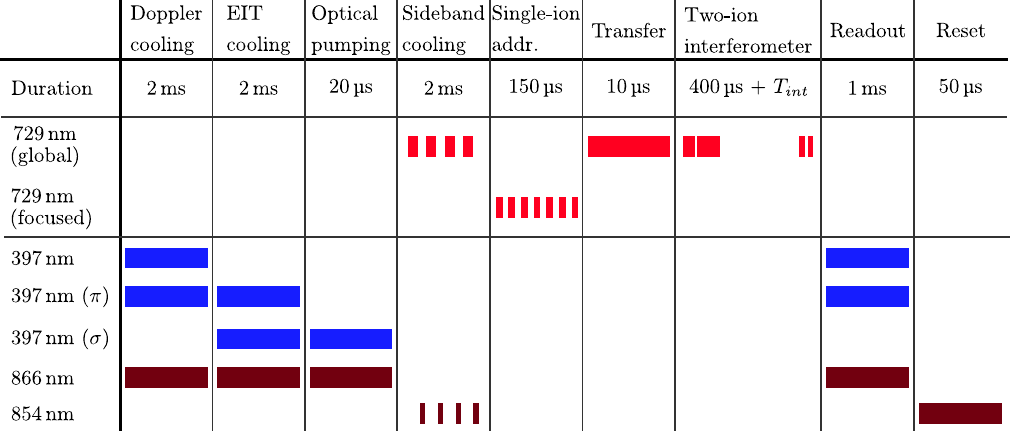}
		\caption{Complete pulse sequence used for a single cycle of the enDFS and ccDFS schemes. 
			Cooling and repumping lasers at \SI{397}{\nano\meter}, \SI{866}{\nano\meter}, and \SI{854}{\nano\meter} are employed alongside the clock laser at \SI{729}{\nano\meter}. 
			The \SI{397}{\nano\meter} cooling laser is applied along three directions, providing $\pi$- and $\sigma$-polarized beams, as well as a beam addressing all motional modes. 
			Active lasers during each step are indicated by horizontal bars, including any applied pulse combinations. 
			Time durations are specified for each step; durations of coherent interactions may vary due to recalibration. 
			The timing shown for the two-ion interferometer corresponds to the enDFS scheme only. Sideband cooling is omitted for the ccDFS scheme. Additional waiting times occur between individual steps and between experimental cycles due to computation and the application of feedback to the clock laser.
		}
		\label{fig:pulse_sequence}
	\end{figure*}

	A detailed description of the individual steps in preparing the ions quantum state and implementation of the interferometer is given.
	The measurement sequence consists of a series of laser pulses generated by acousto-optic modulators (AOM), which are driven by direct-digital synthesizers (DDS) controlled via our experimental control system~\cite{bourdeauducq_artiq_2016}.
	The complete pulse sequence, including time durations and laser wavelengths used in each step, is shown in Fig.~\ref{fig:pulse_sequence}.
	Prior to the interferometric sequence, the ions are cooled sequentially using Doppler, electromagnetically-induced transparency (EIT) and sideband cooling. 
	The initial state \(\ket{\psi_i} = \ket{S_-S_+}\) is prepared by first optically pumping both ions into the \(\ket{S_-S_-}\) state. 
	Subsequently, one ion is coherently transferred to the \(\ket{S_+}\) state via the \(\ket{D_+}\) state using single-ion addressed laser pulses (\(\ket{S_-}\rightarrow\ket{D_+}\) transition) followed by a frequency selective global (\(\ket{D_+}\rightarrow\ket{S_+}\) transition) transfer pulse overlapping with both ions.
	The single-ion addressing step makes use of a passband composite pulse sequence. It is implemented with a focused laser beam on one of the two ions (addressed beam) using seven consecutive pulses \cite{composite_pulses_torosov} on the carrier transition with effective pulse area $\pi$. Especially during the clock measurements, the state preparation fidelity greatly improves by this step, mitigating the effects of beam point fluctuations and crosstalk on the other ion.
	
	The laser pulses for the two-ion interferometer are generated using a DDS connected to a double-pass AOM setup. 
	In the enDFS scheme, the state \(\PhiBellsym\) is prepared using two sideband pulses: a \(\pi/2\) blue sideband (\(\BSB{1}\)) pulse on ion 1, followed by a \(\pi\) red sideband (\(\RSB{2}\)) pulse on ion 2 (see Fig.~\ref{fig:measurement_scheme}). Starting from the electronic ground state, the BSB adds one axial motional quantum, while the RSB removes one~\cite{roos_bell_2004}. 
	The ccDFS scheme is prepared, and both schemes are read out, using two \(\text{CAR}_{(1,2)}(\pi/2)\) pulses. Here, the ions are individually addressed through their transition frequency difference in the weak magnetic field using a laser beam interacting with both ions (global beam).
	
	The four consecutive laser pulses $i$ must maintain a well-defined phase relationship, despite operating at different frequencies $f_i$. The relative phases are kept stable while varying parameters like pulse durations and waiting times by referencing the pulses to a common starting time $t_0$. The phases are dynamically calculated by the measurement system as $\Theta_i = (t_i - t_0) \cdot f_i + \Theta_i$ according to the individual pulse starting time $t_i$. The phases of the first three pulses are set to zero, but could in principle be used to vary the effective readout phase $\Thetaro = \Theta_1 + \Theta_2 + \Theta_3 + \Theta_4$. The implementation of the correlated interrogation scheme employs only two different frequencies but uses the same phase computation. Table~\ref{tab:experimental_parameters} lists couplings strength, pulse durations and detunings used in our experimental runs. These parameters have been calibrated in regular intervals during the clock operation in a series of consecutive scans and vary slightly over the course of the clock measurements. 
	\begin{table*}
		\caption{\label{tab:experimental_parameters} Experimental parameters used for implementation of the two-ion interferometers. The (effective) coupling strength corresponds to the Rabi frequency at the given (zero) detuning. The detuning is specified as the difference $f_0 - f_L$ of the laser on resonance with a frequency $f_L$ against the virtual unperturbed \clocktrans transition frequency $f_0$, but can vary on the \SI{}{\kilo\hertz} level due to ac-Stark shifts.}
		\begin{ruledtabular}
			\begin{tabular}{l|ccccc}
				Pulse & Transition & Coupling strength & Effective coupling strength & Pulse duration  & Detuning\\
				BSB$_1$($\pi/2$) & $\ket{S, m_j = -1/2} \rightarrow \ket{D, m_j = -1/2}$ & $2 \pi \cdot$ \SI{2}{\kilo\hertz} & $2 \pi \cdot$ \SI{85}{\kilo\hertz} & \SI{120}{\micro\second} & \SI{+3.4}{\mega\hertz}\\
				RSB$_2$($\pi$) & $\ket{S, m_j = +1/2} \rightarrow \ket{D, m_j = +1/2}$ & $2 \pi \cdot$ \SI{2}{\kilo\hertz} & $2 \pi \cdot$ \SI{85}{\kilo\hertz} & \SI{240}{\micro\second} & \SI{-3.4}{\mega\hertz}\\
				CAR$_1$($\pi/2$) & $\ket{S, m_j = -1/2} \rightarrow \ket{D, m_j = -1/2}$ & $2 \pi \cdot$ \SI{12.5}{\kilo\hertz} & - & \SI{20}{\micro\second} & \SI{2.0}{\mega\hertz}\\
				CAR$_2$($\pi/2$) & $\ket{S, m_j = +1/2} \rightarrow \ket{D, m_j = +1/2}$ & $2 \pi \cdot$ \SI{12.5}{\kilo\hertz} & - & \SI{20}{\micro\second} & \SI{-2.0}{\mega\hertz}\\
				\hline
			\end{tabular}
		\end{ruledtabular}
	\end{table*}
	The coupling strength is chosen as a trade-off between minimizing off-resonant excitation into different Zeeman states and mitigating the limitations imposed by magnetic field noise, which reduces fidelity.
	
	\subsection{Clock laser and frequency comparison}
	The clock laser setup is based on an extended cavity diode laser (ECDL) locked to a high finesse cavity. We seed another laser diode with the filtered transmission of the cavity \cite{akerman_universal_2015, Nazarova_2008} before amplifying the light in a tapered amplifier. The resulting light has suppressed high frequency noise which benefits the fidelity of the resonant interaction with the sideband transitions. The laser system is further stabilized via a transfer-oscillator lock~\cite{scharnhorst_high-bandwidth_2015} to an ultra-stable reference laser using an optical frequency comb. This ensures traceability of the clock frequency to the reference system and enables long coherence times.
	
	To stabilize the laser to the atomic reference, a two-point sampling method is employed \cite{Peik_2006}. An error signal is derived from ten repetitions of the measurement sequence, using two readout phase settings $\Thetaro = \pm \frac{\pi}{2}$ applied in randomized order. For each setting, the control system computes a parity signal $P_\pm$, and the error signal is defined as $e = \langle P_+ \rangle - \langle P_- \rangle$. This signal is fed into a servo loop with first- (I) and second-order (I$^2$) integrators to estimate the frequency offset. The integrator gains are set to $g_I = 0.3$ and $g_{I^2} = 0.001$, with feedback computed in units of the parity signals fringe width. The laser frequency is then corrected via an rf signal sent to a steering AOM located before the double-path setup and monitored by a frequency counter.
	
\begin{figure*}[t]
		\centering
		\includegraphics[width=\textwidth]{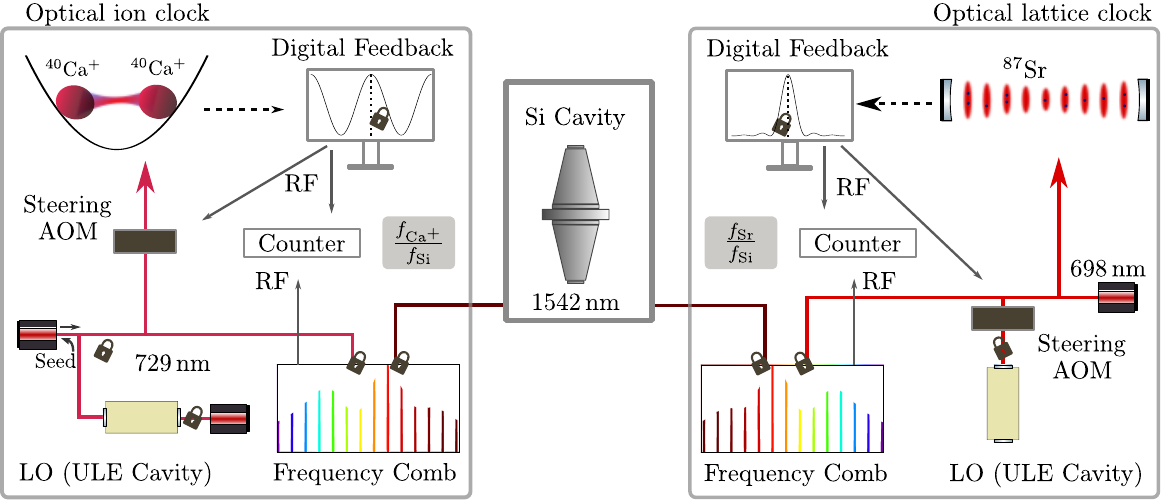}
		\caption{Overview of the optical clock comparison between the trapped-ion and atomic lattice systems. 
			In both experiments, the local oscillator (LO) is locked to the silicon cavity via a frequency comb. 
			The LO consists of a laser tightly stabilized to an ultra-low expansion (ULE) cavity system. In the optical ion clock, the LO comprises two laser diodes and a ULE cavity for locking and filtering. The filtered transmission is used to seed the second laser diode. Associated radiofrequency (rf) signals are tracked by a counter. 
			The LO interrogates the atomic systems, with a digital feedback loop adjusting its frequency based on atomic measurement signals to maintain resonance. 
			Both laboratories independently measure their frequency ratios relative to the silicon cavity.}
		\label{fig:clock_comparison_sketch}
	\end{figure*}

	The optical clock comparison was conducted between two different laboratories on the campus of the Physikalisch-Technische Bundesanstalt in August 2024. Each laboratory is measuring the frequency ratio $\rho_\mathrm{clock, cavity} = \frac{f_\mathrm{clock}}{f_\mathrm{cavity}}$ against a common ultra-stable optical reference system based on a laser stabilized to a Si cavity \cite{matei_si_cavity_2017}. A sketch of the frequency comparison is provided in Fig. \ref{fig:clock_comparison_sketch}. The frequency of the local oscillator (LO) is locked to the reference light via a transfer lock using a frequency comb. All involved phase-locked loops for path-length stabilization, transfer-locked laser systems and the frequency comb's lock to the reference light are tracked by a series of synthonized counter channels, as well as the out-of-loop frequency correction inferred from the clock measurements. The clock's measurement laboratories are providing a set of calculations to derive the ratio 
	\[
	\rho_\mathrm{clock, cavity} = \frac{f_\mathrm{clock}}{f_\mathrm{cavity}}.
	\]
	The ratio between the two clocks \ca, \srneutral is then calculated by 
	\[
	\rho_\mathrm{Ca^+, Sr} = \frac{\rho_\mathrm{Ca^+, cavity}}{\rho_\mathrm{Sr, cavity}}.
	\]
	
	\subsection{Estimation of the Quantum projection noise level}
	The clock measurements were performed over several days. The resulting raw frequency traces were cleaned of events indicating malfunctions in the phase-locked loops. From the mean-subtracted frequency ratio, \(\rho_{\mathrm{rfr}} = \rho_{A,B} - \langle \rho_{A,B} \rangle\), the fractional frequency stability \(\sigma_y\) is calculated as an overlapping Allan deviation (OADEV). Data analysis is carried out using the Python software package \texttt{allantools}.
	
\begin{figure}
		\centering
		\includegraphics[width=0.5\textwidth]{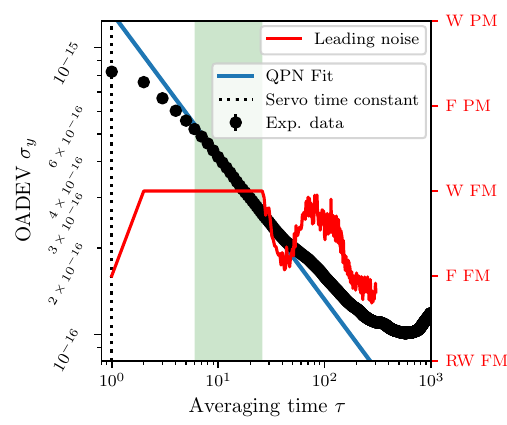}
		\caption{Exemplary sketch showing the fitting of the noise figure from the frequency ratio data for the enDFS scheme at \SI{50}{\milli\second}. Only OADEV-values for averaging times $\tau$ a factor of five larger than the servo time constant are taken into account. An algorithm based on the lag-1 autocorrelation function is used for determination of the leading noise at higher averaging times. The identified noise types at given averaging times are plotted in red. It distinguishes between white phase-modulation (W PM), flicker phase-modulation (F PM), white frequency modulation (W FM), flicker frequency modulation (F FM) and random walk frequency modulation (RW FM). The quantum projection noise is then fitted to the remaining data (green shaded region), and weighted with the statistical error of the OADEV.}
		\label{fig:qpn_fitting_example}
	\end{figure}

	The stability of an atomic clock, characterized by the Allan deviation, can be influenced by various noise processes. For the short averaging times, the noise is a combination of the clock laser noise and the servo response function, characterized by the servo stabilization time constant. In the case of quantum projection noise-limited measurements, the stability is governed by white frequency noise, resulting in an Allan deviation scaling as \(\sigma_y(\tau) = \sigma_y(\SI{1}{\second}) / \sqrt{\tau / \SI{1}{\second}}\)~\cite{rubiola_measurement_2005}. The value \(\sigma_y(\tau = \SI{1}{\second})\) serves as a benchmark for comparing different interrogation schemes. It is both extracted from the measured frequency data and inferred from the theoretical description presented in the main text.
	At long averaging times our clock system is not solely dominated by white frequency noise, requiring the identification of intervals with white frequency noise dominance. This is achieved by analysing the lag-1 autocorrelation function of the frequency ratio data to infer the prevailing noise processes over different averaging times. The employed algorithm is described in~\cite{Riley_frequency_analysis_2008, riley_handbook_2008}. We evaluate the leading noise type by averaging the autocorrelation across all overlapping segments of the data trace. Reliable estimation of the autocorrelation function is possible up to averaging times of approximately \(1/30\) of the total data length. Only data identified as white frequency noise-dominated is considered for stability analysis.
	Additionally, the impact of the servo loop is taken into account, which effectively acts as a low-pass filter on the measurement data. To avoid biasing the stability estimate, only averaging times at least a factor of five longer than the servo time constant are included. This ensures a robust and unbiased estimate of the fractional frequency stability.
	Figure~\ref{fig:qpn_fitting_example} illustrates the analysis for frequency data obtained with an interrogation time of \SI{50}{\milli\second}.
	
\end{document}